# A Reconfigurable Relay for Polarization Encoded QKD Networks

Jing Wang and Bernardo A. Huberman
CableLabs

## Abstract

We propose a method for reconfiguring a relay node for polarization encoded quantum key distribution (QKD) networks. The relay can be switched between trusted and untrusted modes to adapt to different network conditions, relay distances, and security requirements. This not only extends the distance over which a QKD network operates but also enables point-to-multipoint (P2MP) network topologies. The proposed architecture centralizes the expensive and delicate single-photon detectors (SPDs) at the relay node with eased maintenance and cooling while simplifying each user node so that it only needs commercially available devices for low-cost qubit preparation.

# Introduction

State-of-the-art QKD technologies suffer from a trade-off between transmission distance and key rates, therefore necessitating relays as the only way to extend the QKD coverage. While a trusted relay allows for simple implementation and long transmission distance, it becomes a security problem since it knows all the keys. An untrusted relay on the other hand, eliminates the security issues but suffers from a high requirement of detection efficiency and short distance propagation. So far as we know, neither trusted nor untrusted relaying can handle all network conditions independently, and yet there has been no effort to combine the advantages of both technologies.

Since the security of cryptographic algorithms is based on the computational complexity of intractable mathematical problems, up-to-date algorithms are only secure against the state-of-the-art computational power and will always need updates when the growth of computational power exceeds the complexity of those problems. Even post-quantum cryptographic algorithms are only safe against quantum computers with a certain number of qubits, which is not achievable in the near future, but redesigning algorithms is always required as quantum computers keep evolving. On the other hand, quantum key distribution (QKD) offers a disruptive technology to end this competition between computational power and algorithm complexity. It offers information-theoretic security, which is guaranteed by quantum mechanics, i.e. keys cannot be broken even if the adversary has unlimited computing power. Different from the software implementation of cryptographic algorithms, QKD technologies require specific hardware and suffer from a limited distance, low key rate, and high system cost in deployment.

In quantum communication, single-photon pulses are used to carry qubits to avoid photon-number-splitting (PNS) attacks. The absorption of photons in fibers makes the channel loss increase exponentially with distance. Most photons are lost during propagation and only a small portion of them arrives at the receiver. Therefore, the final key rate reduces exponentially with the channel loss and there is a tradeoff between the distance and secure key rate. For example, with fiber attenuation of 0.2 dB/km, after traversing 1000-km fiber (200-dB loss) only 0.3 photons can be detected per centenary even if a 10-GHz photon source was used. The current distance record of QKD in optical fibers is 500 km, reported by Toshiba, whereas the key rate is orders of magnitude lower than practical requirements. In real applications, QKD links in optical fibers are limited to ~100 km without a relay.

An ideal way to extend the distance is to use quantum repeaters, but the three building blocks of a quantum repeater, entanglement swapping, entanglement purification, and quantum memories, are all far away from practical implementations. There have been demonstrations of entanglement swapping over 1000 km, but entanglement purification and qubit storage are still challenged, especially in terms of storage time and retrieval efficiency of qubits. Therefore, the use of relay nodes seems to be the only choice to extend the QKD distance.

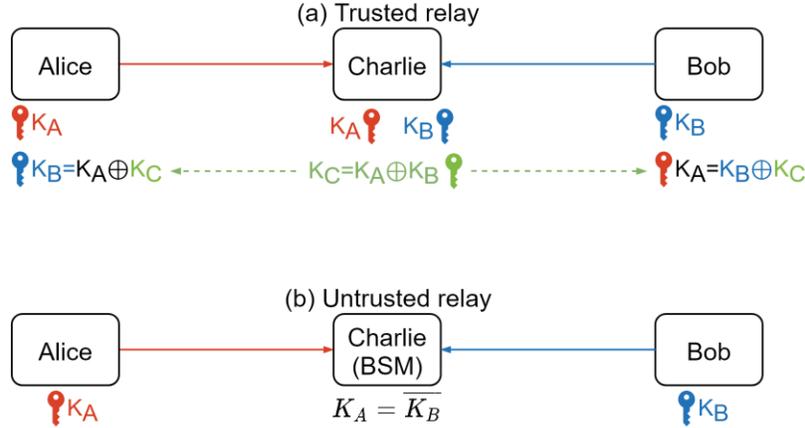

Fig. 1. Operation principles of (a) trusted relay and (b) untrusted relay based on measurement-device-independent (MDI) QKD.

The operation principles of a trusted relay are shown in Fig. 1(a). The relay node, Charlie, first performs QKD with Alice and Bob and obtains keys of $K_A$ and $K_B$, respectively, then announces the bitwise parity-check of $K_A$ and $K_B$, $K_C=K_A\oplus K_B$. Since $K_A$ and $K_B$ are independent bit strings, their bitwise parity is a uniformly random bit string, which does not reveal any information about the keys. With the help of $K_C$, both Alice and Bob can infer each other's keys using the fact that $K_A\oplus(K_A\oplus K_B)=K_B$ and $K_B\oplus(K_A\oplus K_B)=K_A$. Since Charlie holds all the keys, any access to the relay node gives an adversary complete knowledge of the keys. In Fig. 1(a), the directions of QKD links are chosen to be from Alice/Bob to Charlie. This is because a QKD channel is unidirectional with unbalanced costs of transmitter and receiver. A quantum transmitter only consists of commercially available devices, e.g., lasers, modulators; whereas a quantum receiver consists of expansive and delicate single-photon detectors (SPDs), which require cooling and triggering. The channel direction from Alice/Bob to Charlie centralizes SPDs in the relay node, where the cost of quantum receivers is shared by multiple users. This strategy not only reduces the overall system cost but also eases the cooling and synchronization of SPDs.

An untrusted relay can be implemented by either measurement-device-independent (MDI) QKD or using entangled photon pairs. We point out that entanglement-based protocols require an expensive photon source to generate entangled photon pairs (EPR) using nonlinear optics. Here we only focus on MDI-QKD, shown in Fig.1(b). Both Alice and Bob randomly and independently prepare qubits and send them to Charlie. Charlie performs Bell state measurements (BSM) and makes a public announcement of the results of successful BSMs. Since MDI-QKD removes all loopholes at the detection side, Charlie has no information about the qubits sent from Alice and Bob. But for successful BSM events, the qubits from Alice and Bob are entangled, e.g., complementary to each other, so Alice and Bob can infer each other's keys. Instead of generating entangled photons like EPR protocols, MDI-QKD post-selects entangled photons by BSM performed by Charlie. This is why MDI-QKD is equivalent to a *time-reversed EPR protocol*. In this way, Charlie serves as an untrusted relay and could even be an eavesdropper itself. Since the key generation in MDI-QKD is based on coincident detection events of photons from two users, the secure key rate has a quadratic dependence on the detection efficiencies of SPDs. Thus, MDI-QKD has higher requirements on detection efficiency and suffers from a shorter distance limit and lower key rate than a trusted relay.

Most prepare-and-measure QKD protocols are limited to point-to-point (P2P) connections and cannot handle complicated network topologies. Relay nodes not only extend the coverage of QKD links but also help to handle point-to-multipoint (P2MP) quantum networks. They are intrinsically desirable for metro and access networks with mesh, star, or tree topologies where the relay nodes are located at hubs where complicated and expensive quantum receivers are centralized and shared by multiple users. Other nodes only need low-cost commercial devices for qubit preparation. To add a new node, only lasers and modulators are needed and there is no upgrade at the relay node. Low hardware requirements and small upgrade costs make relaying networks scalable for a large number of users.

In summary, a trusted relay features simple implementation, long distances, and high key rate, and also offers good compatibility to different QKD protocols, i.e., diverse QKD protocols could be used for the links from users to the relay node. On the other hand, an untrusted relay provides higher security by closing all loopholes at the detector side but suffers from a shorter distance and lower key rate. Instead of competing, the two relay technologies should complement each other to accommodate different network conditions. Comparing Fig. 2(a) and (b), both relay technologies have quantum links from users to the relay node, but the difference is that for an untrusted relay, Alice and Bob have to send photons to Charlie at the same time. Furthermore, the configurations of quantum receivers at the relay node are quite similar for both trusted and untrusted relays, which motivates the research of a reconfigurable receiver for both technologies.

To date, neither trusted nor untrusted relaying can independently handle all network conditions and there has been no effort to combine the advantages of both technologies. We thus propose a reconfigurable relay node that can switch between trusted and untrusted modes without significantly increasing the system's cost. The proposed relay node is based on a simple modification of the existing BB84 receiver and makes both relay technologies complement each other to adapt to different network conditions.

## A Configurable Quantum Relay

Most QKD protocols are limited to P2P connections and thus cannot handle complicated network topologies. Relay nodes not only extend QKD distance but also help to handle P2MP networks. In Fig. 2(a), without relaying, each pair of nodes needs a dedicated QKD link for key exchange. From the deployment perspective, this architecture is highly impractical and not scalable since almost every node has to be equipped with an expensive quantum receiver and the number of QKD links increases quadratically with the number of nodes. From the security perspective, however, it is robust since any compromised node does not breach the security of others.

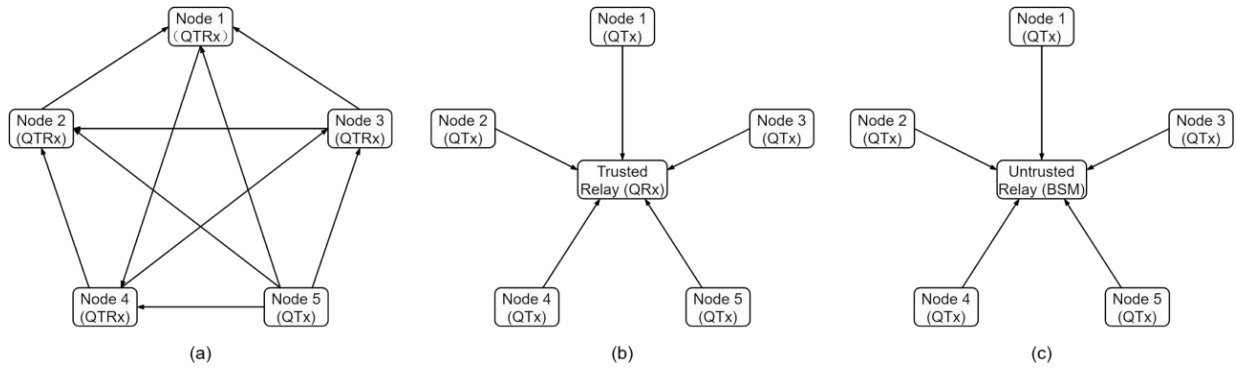

Fig. 2. Quantum key distribution (QKD) networks. (a) A dedicated QKD link is required for each pair of nodes. (b) A trusted relay network simplifies the topology and reduces the number of QKD links, but the relay node knows all keys. (c) An untrusted relay network keeps the simplicity of trusted relay architecture but removes the vulnerability of the relay node.

Fig. 2(b) shows a trusted relay network. Instead of deploying a QKD link between every two nodes, every node exchanges keys with the relay node, and any two nodes can infer each other's keys via the relay node. Given the imbalanced hardware requirements of quantum transmitter and receiver, only the relay node is equipped with an expensive quantum receiver, which is shared by other nodes. All other nodes only need a quantum transmitter consisting of low-cost devices for qubit preparation. By reducing the number of QKD links to the number of users, this architecture not only saves the system cost and simplifies the network topology, but also makes a scalable QKD network possible. To add a new user, only one quantum link is needed, and only commercially available devices are required for the new user. At any given time, only one user is allowed to send qubits to the relay node. Once the relay node exchanges key with all users, it can relay keys between any two users by announcing the parity-check of their keys. The only drawback of this network is the security of the relay node. Once it is compromised, the whole network is breached.

Fig. 2(c) shows an untrusted relay network based on MDI-QKD, which keeps the simplicity of the trusted relay while eliminating its vulnerability. By replacing the trusted relay with an untrusted one, any two users can exchange keys via the relay node without worrying about the leakage of keys. Any two nodes independently prepare their quantum states and send them to the relay node. The relay node performs BSMs on the incoming photons and announces the results of successful BSMs. The relay node is used as a public detection server and does not need to be trusted or certified. Notice that certification of a quantum detector has been a major hurdle to the standardization of QKD since manufacturers can leave backdoors in detectors and steal information. MDI-QKD solves this problem by eliminating the necessity of detector certification. An untrusted relay network is more resilient than a trusted relay network since any compromised node does not breach the security of others.

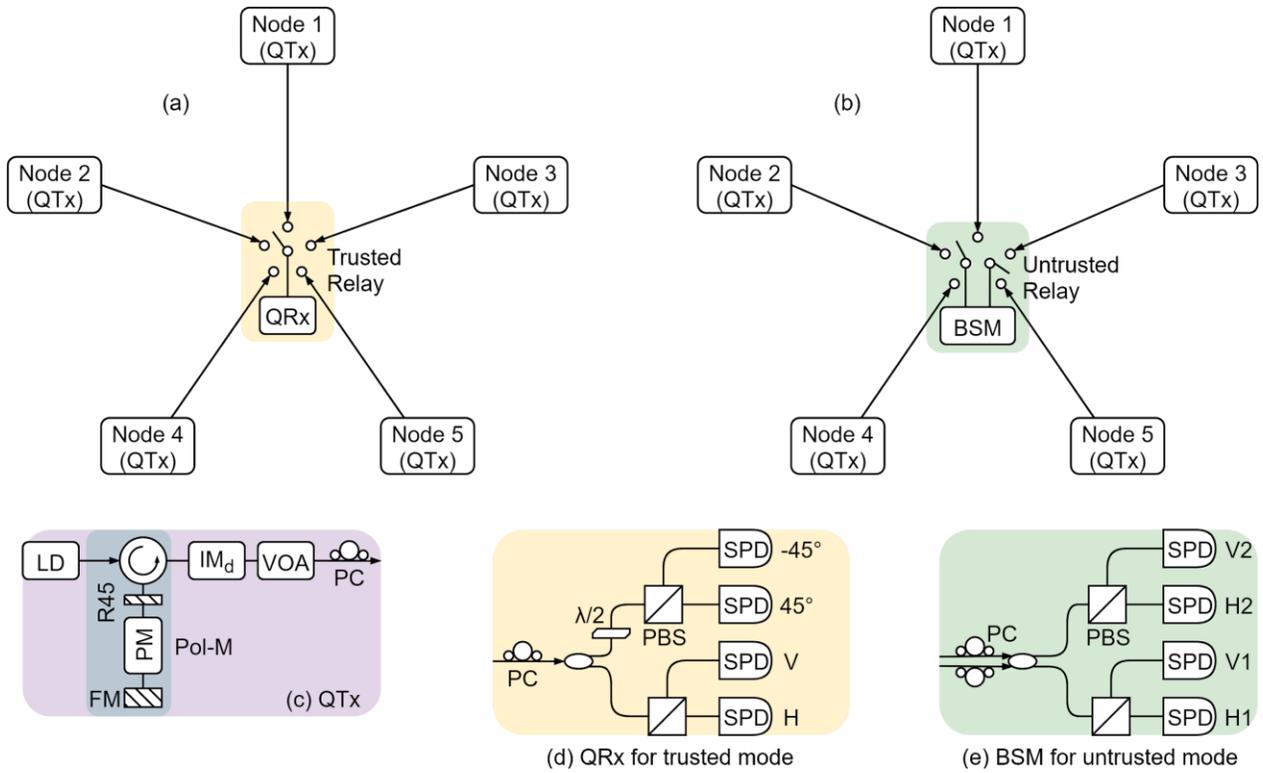

Fig. 3. The proposed polarization encoded QKD network with a reconfigurable relay node. (a) For the trusted mode, the BB84 protocol is used in QKD links from each user to the relay node. (b) In the untrusted mode, polarization-encoding MDI-QKD protocol is used, where two users send single photons to the relay node for BSM. (c) Quantum transmitter at each user node. (d) Quantum receiver at the trusted relay node for BB84 protocol. (e) BSM setup at the untrusted relay node for MDI-QKD.

Fig. 3 shows the proposed QKD network with a reconfigurable relay node. Fig. 3(a) shows the trusted mode, where a polarization encoded BB84 protocol is used for the QKD link from each user to the relay node. At any given time, the optical switch at the relay node only allows one user to send qubits to the relay node. Each user node is equipped with a quantum transmitter shown in Fig. 3(c), and the relay node is equipped with a quantum receiver as shown in Fig. 3(d). Fig. 3(b) shows the untrusted mode, where a polarization-encoding MDI-QKD protocol is used. The optical switch allows two users to send qubits to the relay node at the same time. The quantum transmitter setup at each user node remains the same, but the BB84 receiver at the relay node is reconfigured to a BSM setup, as shown in Fig. 4(e). The second input port of a 2x2 beam splitter is used for two-photon interference, whereas in Fig. 4(d) only one input is used. Another modification is that the half-wave plate in Fig. 4(d) for diagonal basis detection was removed, so all four single-photon detectors (SPDs) only detect rectilinear polarizations.

**Reconfigurable Setup**

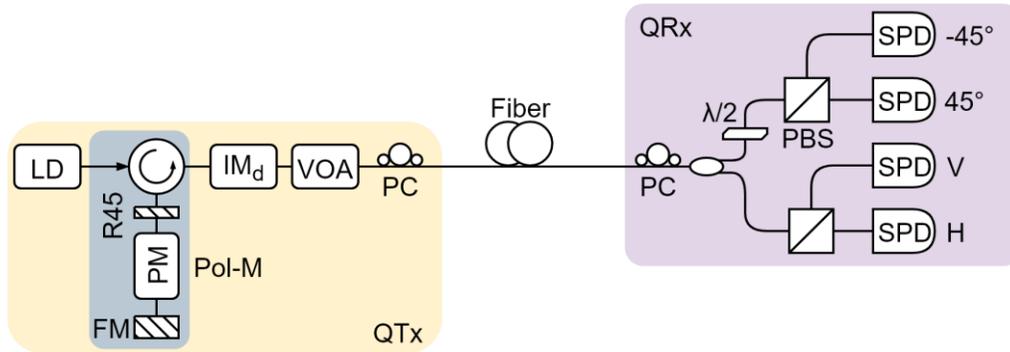

Fig. 4. QKD setup for the trusted relay mode.

Fig. 4 shows the QKD setup for the trusted relay, which is the well-known polarization-encoding BB84 setup. Each user only has a quantum transmitter. A quantum receiver is equipped at the relay node and shared by multiple users. At a user node, the polarization modulator (Pol-M) consists of a circulator, a phase modulator (PM), and a Faraday mirror. The circulator separates the input and output optical pulses. The PM has polarization-maintaining input fiber (PMF) and standard single-mode output fiber. The slow axis of the input PMF is aligned at 45° (R45) to the optical axes of the modulator waveguide. A horizontally polarized input light parallel to the slow axis of the PMF is split into two components, aligned to the fast/slow axis of the waveguide respectively. With a drive voltage applied on the PM, a phase shift is introduced between the two components, resulting in a polarization rotation. To compensate for the polarization mode dispersion and temperature variation of the waveguide, a Faraday mirror is used to rotate the polarization by 90°, so the optical pulse travels through the PM twice with orthogonal polarizations. The voltage signal applied to PM is a stream of pulses synchronized with the propagation of optical pulses so that each optical pulse is modulated only once in the roundtrip.

With the Pol-M, one user randomly prepares photons of four polarization states in two conjugate bases, rectilinear (0°, 90°) and diagonal (±45°), and send them to the relay node. To close the loophole of the photon-number-split attack, an intensity modulator ($IM_d$) is used to adjust the photon number per pulse for decoy-state generation. At the relay node, a 50:50 beam splitter randomly selects the measurement bases. In the lower branch, a polarization beam splitter (PBS) projects the incident photon to horizontal or vertical polarization of the rectilinear basis. In the upper branch, a half-wave plate rotates the polarization of the incident photon by 45° and projects it to the diagonal basis.

Fig. 5 shows the QKD setup for the untrusted relay mode, which is a polarization-encoding MDI-QKD setup. The transmitter scheme at each user remains the same as Fig. 4. Two user nodes, Alice and Bob, independently and randomly prepare photons of four polarization states in two conjugate bases, rectilinear (0°, 90°) and diagonal (±45°), and simultaneously send these photons to the relay node Charlie for BSM. At the relay node, two photons from Alice and Bob interfere at a 50:50 beam splitter and are projected to horizontal (H) and vertical (V) polarization by two PBS. The incoming photons are then detected by four SPDs and registered by a time-interval analyzer (TIA). Compared with Fig. 5, the only two differences in Fig. 5 include 1) a second input port of the 2x2 beam splitter is used for two-photon interference; 2) the half-wave plate in Fig. 5 for diagonal basis measurement is removed and only rectilinear polarizations are detected.

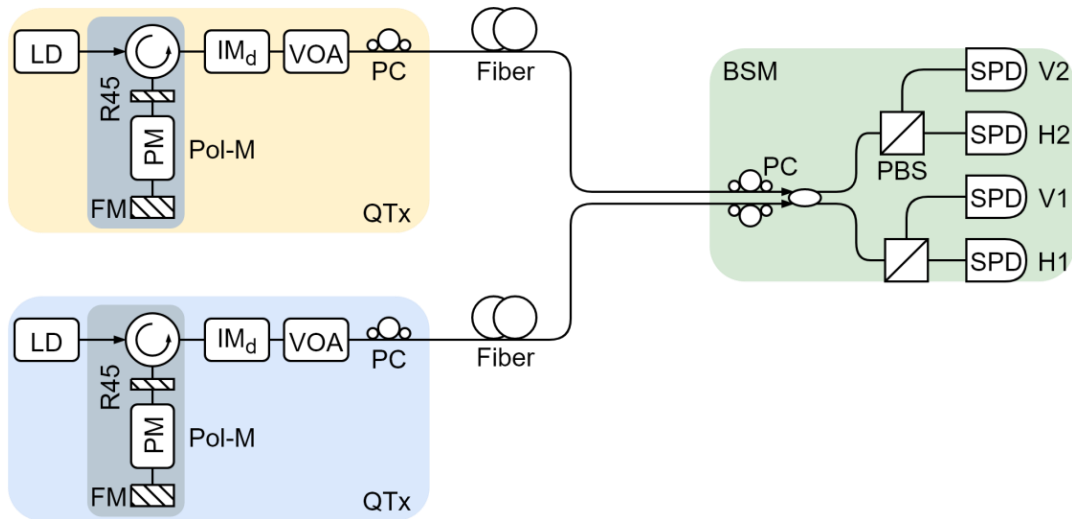

Fig. 5. QKD setup for the untrusted relay mode.

**Operation Principles**

Fig. 6 shows the operation flowchart of the trusted relay. Two user nodes, Alice and Bob, first exchange keys with the relay node Charlie via the BB84 protocol respectively. Then Charlie makes a public announcement of the parity-check of their keys so Alice and Bob can infer each other's keys. The BB84 protocol that each user practices with Charlie is shown in the grey inset. Its operation principles are shown in Table 1. Alice first randomly selects rectilinear or diagonal bases, then randomly chooses bits, and encodes these bits onto the four polarization states (0°, 90°, ±45°) in two conjugate bases. Charlie receives these photons and randomly selects the measurement bases. If a photon is prepared and measured in the same basis, the qubit it carries can be retrieved without error. If it is prepared and measured in different bases, the measurement result is random since the two bases are conjugate to each other. During the basis reconciliation, Alice and Charlie discuss their basis choices via an authenticated public channel. During the key sifting, those bits that are prepared and measured in incompatible bases will be discarded (grey in Table 1) and the rest will be kept as the raw keys.

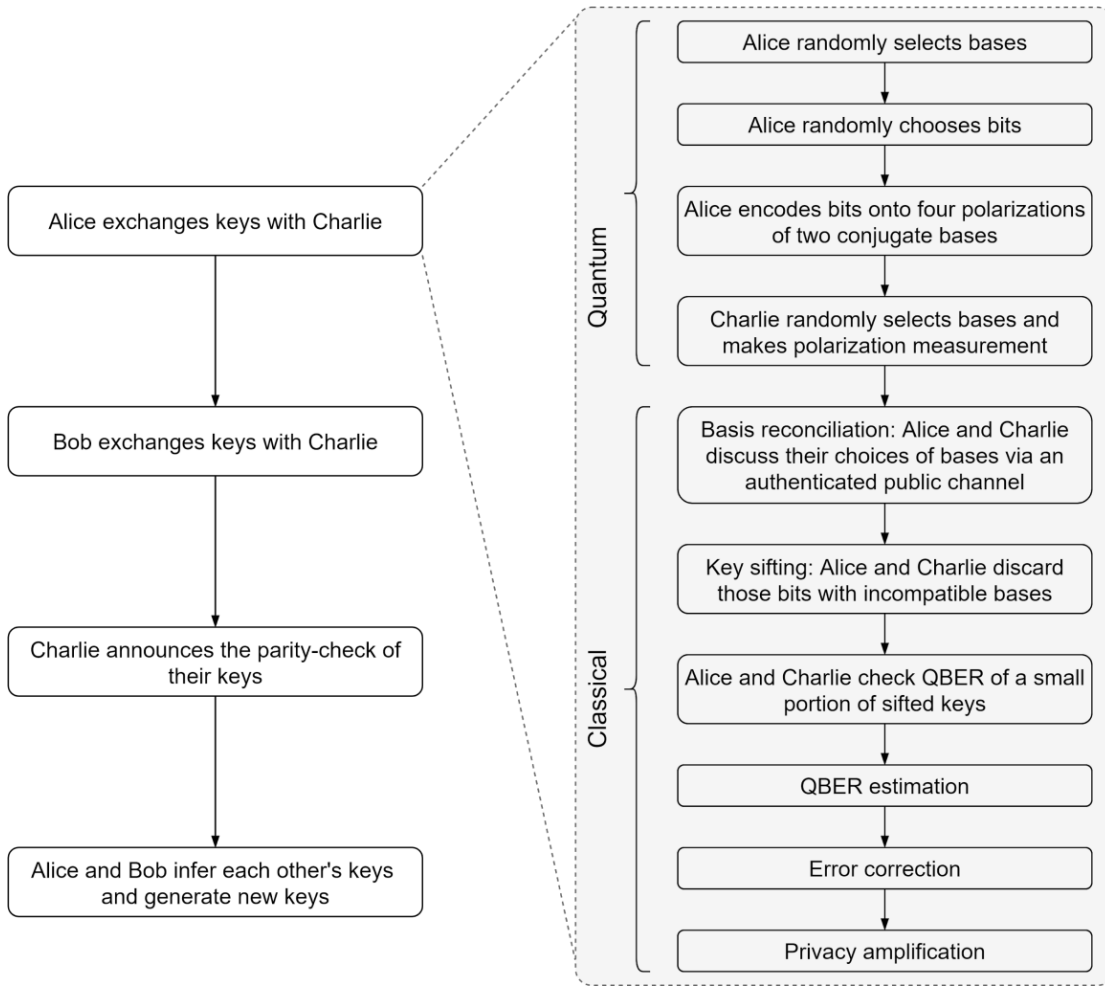

Fig. 6. Operation flowchart of trusted relay

Table 1. Operation principles of BB84 protocol

| Alice bases | + | + | x | + | x | x | + | x | + | + | x | x |
|---|---|---|---|---|---|---|---|---|---|---|---|---|
| Alice bits | 1 | 0 | 0 | 1 | 0 | 1 | 0 | 1 | 0 | 1 | 1 | 0 |
| Polarization | ↑ | → | ↗ | ↑ | ↗ | ↘ | → | ↘ | → | ↑ | ↘ | ↗ |
| Charlie bases | x | + | x | x | + | x | + | + | x | + | x | + |
| Charlie bits | ? | 0 | 0 | ? | ? | 1 | 0 | ? | ? | 1 | 1 | ? |

Fig. 7 shows the operation flowchart of the untrusted relay. Two user nodes, Alice and Bob, independently and randomly select transmitting bases and encode random qubits onto four polarization states of two conjugate bases. They send photons to the relay node Charlie at the same time for BSM. Charlie performs BSM and announces the results of successful BSMs. A successful BSM is defined as coincidence detection events at two orthogonal SPDs. In Fig. 5, if the two SPDs connect to the same PBS, H1+V1 or H2+V2, the received state is a triplet state $|\psi^+\rangle = (|HV\rangle + |VH\rangle)/\sqrt{2}$. If the two SPDs connect to different PBS, H1+V2 or H2+V1, the received state is a singlet state $|\psi^-\rangle = (|HV\rangle - |VH\rangle)/\sqrt{2}$. Since these are two out of four states of an entangled photon pair, this measurement is also called partial BSM. All other detection events, such as detections at only one SPD and coincident detections at two SPDs with parallel polarization (H1+H2, V1+V2), are ignored. The BSM post-selects entangled photons from Alice and Bob but gives no information to Charlie about their qubits. Alice and Bob only keep those bits that result in successful BSM events as raw keys. All other bits are discarded. During the basis reconciliation, Alice and Bob discuss their choices of bases via an authenticated public channel. During the key sifting, Alice and Bob only keep bits that they select the same bases and discard the others that they use incompatible bases. After that, error correction and privacy amplification are performed for key distillation.

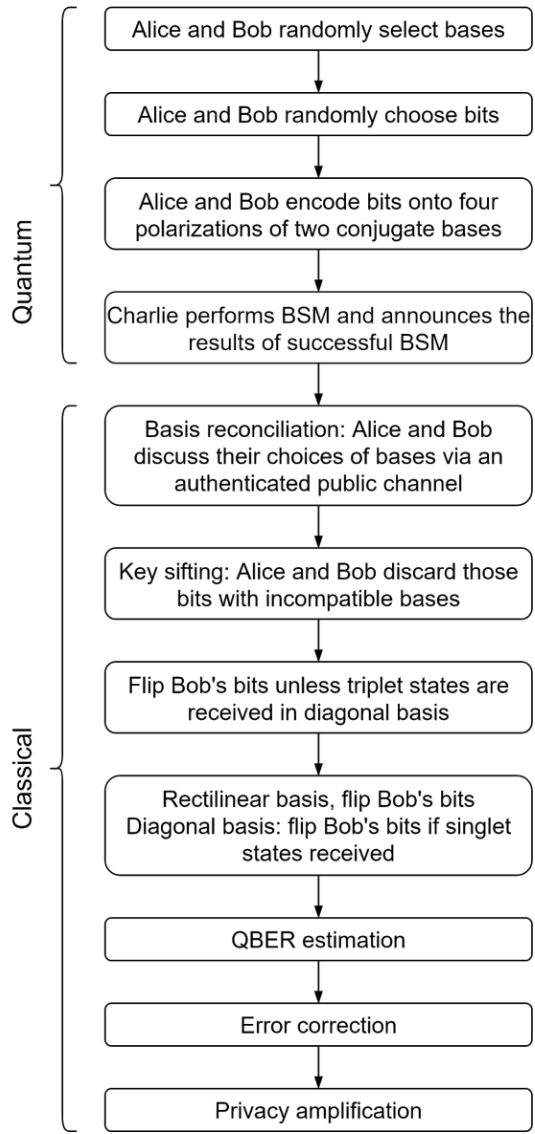

Fig. 7. Operation flowchart of untrusted relay

Table 2 explains the operation principle of polarization-encoding MDI-QKD. All cases when Alice and Bob use incompatible bases are discarded during the key sifting. According to the photon bunching effect of Hong-Ou-Mandel (HOM) interference, when two indistinguishable photons with identical frequency and polarization enter a 50:50 interference beam splitter, they always exit via the same output. Consider the rectilinear basis first. If Alice and Bob both send photons polarized at 0°, the two photons must exit the beam splitter together and both hit H1 or H2. If Alice and Bob both prepare photons at 90°, the two photons will both hit V1 or V2. In these cases, only one SPD receives photons and no successful BSM is made. If Alice and Bob prepare photons at orthogonal polarizations, the two photons may or may not exit the beam splitter together. If they do, they will hit H1+V1 or H2+V2, corresponding to a triplet state (red in Table 2); if they do not, they will hit H1+V2 or H2+V1, corresponding to a singlet state (blue).

Table 2. Operation principles of polarization-encoding MDI-QKD

| Bob \ Alice | 0° (bit 0) | 90° (bit 1) | -45° (bit 0) | 45° (bit 1) |
|---|---|---|---|---|
| 0° (bit 0) | 2xH1<br>2xH2 | H1+V1 (triplet)<br>H2+V2 (triplet)<br>H1+V2 (singlet)<br>V1+H2 (singlet) | Incompatible bases<br>Discarded during key sifting | |
| 90° (bit 1) | H1+V1 (triplet)<br>H2+V2 (triplet)<br>H1+V2 (singlet)<br>V1+H2 (singlet) | 2xV1<br>2xV2 | | |
| -45° (bit 0) | Incompatible bases<br>Discarded during key sifting | | 2xH1, 2xV1<br>H1+V1 (triplet)<br>2xH2, 2xV2<br>H2+V2 (triplet) | 2xH1, 2xV1<br>2xH2, 2xV2<br>H1+H2, V1+V2<br>H1+V2 (singlet)<br>V1+H2 (singlet) |
| 45° (bit 1) | | | 2xH1, 2xV1<br>2xH2, 2xV2<br>H1+H2, V1+V2<br>H1+V2 (singlet)<br>V1+H2 (singlet) | 2xH1, 2xV1<br>H1+V1 (triplet)<br>2xH2, 2xV2<br>H2+V2 (triplet) |

Table 3. Bit flip after the key exchange

| Tx Basis \ BSM results | Singlet state (H1+V2, V1+H2) | Triplet state (H1+V1, H2+V2) |
|---|---|---|
| Rectilinear | Flip | Flip |
| Diagonal | Flip | No flip |

Now consider the diagonal basis. If Alice and Bob prepare photons with the same polarization, the two photons must exit the splitter together. Taking port 1 as an example, there are three possible outcomes, both to H1, both to V1, one to H1 and one to V1. The last case is a triplet state (red). If Alice and Bob prepare photons with orthogonal polarizations, -45° and 45°, the two photons may or may not exit the beam splitter together. If they do, according to the photon bunching effect, they will always reach the same detector, i.e. both to H1 or V1 if exiting via port 1, both to H2 or V2 if exiting via port 2. Therefore, there is no coincident detection event of two SPDs if the two photons exit the beam splitter together. If the two photons exit the splitter separately, there are four possible outcomes, H1+H2, V1+V2, H1+V2, V1+H2. The last two are singlet states (blue). In Table 2, for the rectilinear basis, the occurrence of singlet and triplet states is overlapped, i.e. photons from Alice and Bob have orthogonal polarizations and their corresponding bits are complementary. For the diagonal basis, however, the occurrence of singlet and triplet states are exclusive. Only the triplet state is possible if Alice and Bob prepare the same polarization; only singlet state possible if they prepare orthogonal polarizations.

Table 3 shows the required bit flip after key exchange. For the rectilinear basis, successful BSMs only occur when Alice and Bob send photons in orthogonal polarizations. Either one of them needs to flip the bit. For the diagonal basis, bit flip is still required for singlet states, but no bit flip is needed if a triplet state is received. In summary, either Alice or Bob needs to flip his/her bit unless they use the diagonal basis and Charlie receives a triplet state. In the untrusted mode, the relay node serves as a public BSM server to help any two user nodes exchange keys. There is no leakage of key information even if the relay node is controlled by an eavesdropper.

## Conclusions

In this paper, we proposed a system and method for a reconfigurable relay node for polarization encoding QKD networks, which can be switched between trusted and untrusted modes to adapt to different network conditions, relay distance, and security requirements. The proposed relay node not only circumvents the tradeoff between distance and key rate of existing QKD networks but also enables P2MP network topologies. The resulting network architecture centralizes the expensive and delicate SPDs at the relay node to facilitate easy maintenance and cooling. Each user node is simplified and only needs low-cost commercial devices for qubit preparation. By exploiting the similarity between the quantum receiver of the BB84 protocol and the BSM setup of MDI-QKD, our reconfigurable design enables two different relaying modes without significant modification or cost increase. Moreover, it reduces the system cost since the expensive quantum receiver is shared by multiple users.

In the trusted mode, the relay node first exchanges keys with each user before relaying keys among them, so the relay node becomes a security loophole since it knows all the keys. In the untrusted relay, the relay node facilitates key exchange between any two users as a public server for BSM but has no access to their keys. Since MDI-QKD relies on coincident detection events of two photons, it has higher requirements of detection efficiency of SPDs and is limited to a shorter distance than regular prepare-and-measure protocols. The two relaying modes have to be chosen according to the network conditions and security requirements.

**Acronym**

BSM           bell state measurement

EPR           entangled photon pair

HOM           Hong-Ou-Mandel

IM            intensity modulator

| | |
|---|---|
| MDI | measure-device-independent |
| P2P | point-to-point |
| P2MP | point-to-multipoint |
| PBS | polarization beam splitter |
| PM | phase modulator |
| PMF | polarization maintaining fiber |
| PNS | photon number splitting |
| Pol-M | polarization modulator |
| QKD | quantum key distribution |
| RSA | Ron Rivest, Adi Shamir and Leonard Adleman |
| SPD | single photon detector |
| TIA | time-interval analyzer |